\documentclass[11pt,a4paper]{article}
\usepackage{jheppub}
\usepackage{amsmath,amssymb,amsfonts}
\usepackage{graphicx,graphics}
\usepackage{graphics}
\usepackage{url}
\usepackage{slashed}
\usepackage{color}
\usepackage{mathtools}
\usepackage[utf8]{inputenc}
\usepackage{subcaption}

\newcommand{\msbar}{\overline{\mathrm{MS}}}
\newcommand{\renorm}{\mu_{\text{ren}}}
\newcommand{\fact}{\mu_{\text{fact}}}
\newcommand{\frag}{\mu_{\text{frag}}}
\newcommand{\dd}{\mathrm{d}}

\title{Dimuon production in neutrino-nucleus collisions at next-to-next-to-leading order in perturbative QCD}

\affiliation{University of Jyväskylä, Department of Physics, P.O. Box 35, FI-40014 University of Jyväskylä, Finland}
\affiliation{Helsinki Institute of Physics, P.O. Box 64, FI-00014 University of Helsinki, Finland}

\emailAdd{ilkka.m.helenius@jyu.fi}
\emailAdd{hannu.paukkunen@jyu.fi}
\emailAdd{sami.a.yrjanheikki@jyu.fi}

\abstract{
Charm production in charged-current neutrino-nucleus deep-inelastic scattering (DIS), measured through dimuon final states, remains an important constraint of strangeness in global analyses of parton distribution functions (PDFs). This process has traditionally favored a smaller strange-quark PDF at small momentum fractions $x$ than what the LHC heavy-gauge boson data have indicated. Here, we present a self-contained next-to-next-to-leading-order (NNLO) perturbative QCD calculation of dimuon production in neutrino-nucleus DIS based on semi-inclusive DIS (SIDIS). This process has been previously computed at NNLO through fully inclusive charm production. We discuss the shortcomings of this approach and how they are addressed in the SIDIS picture. We study the perturbative convergence and explore new heavy-quark production channels that become available at NNLO. We find that the NNLO corrections significantly reduce the scale uncertainties at large values of $x$ where the cross sections are enhanced by the NNLO corrections. At small $x$, the NNLO corrections tend to be negative instead, which alleviate the tension between the dimuon and LHC data.
}


\begin{document}

\author{Ilkka Helenius,}
\author{Hannu Paukkunen and}
\author{Sami Yrjänheikki}

\maketitle

\section{Introduction}
\label{sec:Intro}

Parton distribution functions (PDFs) \cite{Kovarik:2019xvh,Ethier:2020way,Klasen:2023uqj} and fragmentation functions (FFs) \cite{Metz:2016swz} are important non-perturbative inputs in applications of perturbative Quantum Chromodynamics (QCD) within the framework of collinear factorization \cite{Collins:1989gx}. While there has been significant advances in the first-principle calculations of PDFs \cite{Lin:2017snn,Constantinou:2020hdm}, the traditional way to determine PDFs from experimental data in global analyses is still the leading one. As of today, free-proton PDF analyses are routinely done at NNLO in perturbative QCD \cite{Alekhin:2017kpj,Hou:2019efy,Bailey:2020ooq,NNPDF:2021njg,ATLAS:2021vod}, with N$^3$LO fits already on the horizon \cite{McGowan:2022nag,NNPDF:2024nan,Ablat:2025gdb,Cridge:2025oel}. Nuclear PDF fits, however, are still mostly considered at NLO \cite{Duwentaster:2022kpv,Eskola:2021nhw,AbdulKhalek:2022fyi}, although a few NNLO fits already exist \cite{AbdulKhalek:2019mzd,Khanpour:2020zyu,Helenius:2021tof}. On the one hand, this disparity has been due to a lack of precise data which would necessitate the inclusion of NNLO corrections. On the other hand, the applications of nuclear PDFs e.g. in heavy-ion collisions typically involve modeling beyond collinear factorization where the perturbative accuracy does not play such a central role.  However, with the promise of new precise data, particularly from new efforts at the Large Hadron Collider (LHC), the upcoming High-Luminosity LHC (HL-LHC) \cite{Cruz-Martinez:2023sdv,Ariga:2025qup,FASER:2019dxq,SNDLHC:2022ihg,Feng:2022inv,Alekhin:2015byh,SHiP:2025ows,FPF:2025bor}, and from the Electron-Ion Collider (EIC) \cite{AbdulKhalek:2022hcn}, NNLO nuclear PDFs will become increasingly relevant. For electroweak-boson production in p+Pb collisions at the LHC the NNLO corrections already exceed experimental uncertainties at certain kinematics \cite{Helenius:2021tof}.

The aim of this paper is to discuss the calculation of dimuon production in neutrino-nucleus collisions at NNLO precision. In this charged-current process, one muon originates from the beam neutrino and the other one from a heavy-flavor decay providing access to the strange-quark content of the target nucleon. While the available measurements for this process from the CCFR, NuTeV, and NOMAD collaborations \cite{CCFR:1994ikl,NuTeV:2001dfo,NuTeV:2007uwm,CHORUS:2008vjb,NOMAD:2013hbk}, taken with heavy nuclear targets (mainly iron), date back 10 to 30 years, they still constitute an important constraint of strangeness in nuclei and for the possible strange-antistrange asymmetry. More recent constraints include e.g. inclusive and charm-tagged $W$ and $Z$ production in proton-proton collisions \cite{Kusina:2012vh,Bevilacqua:2021ovq,Generet:2025bqx}, as well as semi-inclusive kaon production in charged-lepton-proton collisions \cite{HERMES:2013ztj,Borsa:2017vwy,Sato:2019yez}. There are, however, also challenges when it comes to constraining the strangeness by these processes in global analyses. In $W$ and $Z$ production, the high interaction scale makes it more difficult to disentangle the non-perturbative strange distribution from the perturbatively-generated strange sea. In kaon production, the precision of the kaon FFs proves to be a limiting factor. In addition, the LHC $W$ and $Z$ production data appear to be in tension with the dimuon data: the strange-sea suppression $(s+\bar s)/(\bar u+\bar d)$ determined from neutrino data is around 0.5 \cite{CCFR:1994ikl,NOMAD:2013hbk} at $x \sim 0.02$ at factorization scale $Q \sim 1.6 \, \mathrm{GeV}$, whereas the LHC data prefer an unsuppressed strange sea \cite{ATLAS:2012sjl,ATLAS:2016nqi,Cooper-Sarkar:2018ufj,NNPDF:2017mvq}. There are, however, indications \cite{Faura:2020oom} that the NNLO corrections will reduce these tensions, which makes it important to scrutinize these corrections.

\begin{figure}
\centering
\begin{subfigure}{0.33\textwidth}
\centering
\includegraphics[]{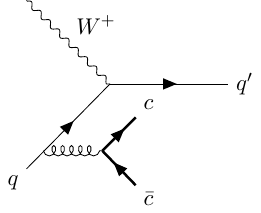}
\caption{}
\label{fig:nnlo_charm_pair}
\end{subfigure}%
\begin{subfigure}{0.33\textwidth}
\centering
\includegraphics[]{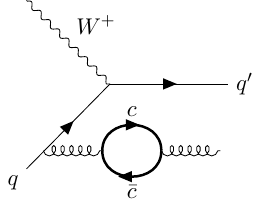}
\caption{}
\label{fig:nnlo_rv_charm_loop}
\end{subfigure}%
\begin{subfigure}{0.33\textwidth}
\centering
\includegraphics[]{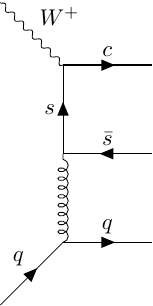}
\caption{}
\label{fig:nnlo_square}
\end{subfigure}
\caption{Examples of NNLO diagrams in neutrino-nucleus DIS.}
\end{figure}

The dimuon production $\nu N\to \mu\mu X$ is typically calculated by relating it to the inclusive charm production $\nu N\to c\mu X$ by a factorization-like formula,
\begin{equation}
\label{eq:charm_factorization}
    \dd\sigma(\nu N\to \mu\mu X)=\mathcal{A}\mathcal{B}_\mu \dd\sigma(\nu N\to c\mu X),
\end{equation}
where the acceptance correction $\mathcal{A} \sim 0.4 \ldots 0.9$ ensures that the decay muon meets the experimentally-imposed energy cut, and $\mathcal{B}_\mu\sim 0.09 \pm 10 \, \%$ is the average semileptonic branching ratio. This is accurate to leading order (LO) in the fixed-flavor number scheme (FFNS) --- a framework in which the charm-quark is not considered as an active parton in the Dokshitzer-Gribov-Lipatov-Altarelli-Parisi (DGLAP) evolution \cite{Gribov:1972ri,Gribov:1972rt,Dokshitzer:1977sg,Altarelli:1977zs}. Beyond LO FFNS eq.~\eqref{eq:charm_factorization} becomes increasingly approximative and just by increasing the perturbative accuracy of $\dd\sigma(\nu N\to c\mu X)$ does not necessarily mean that the overall accuracy would be increased unless the acceptance correction --- taken by default from a fixed-order NLO calculation \cite{Mason:2006qa} --- is also updated to NNLO precision. Furthermore, to consistently account for the heavy-quark mass effects in association with PDFs, an appropriate general-mass variable-flavour-number scheme (GM-VFNS) should be applied in the acceptance calculation \cite{Olness:1987ep,Aivazis:1993kh,Aivazis:1993pi,Kretzer:1997pd,Collins:1998rz,Kramer:2000hn,Tung:2001mv,Kretzer:2003it,Thorne:2008xf,Guzzi:2011ew,Gao:2021fle,Risse:2025smp}. Nevertheless, such a factorized approximation has been applied in the NNLO calculations of the dimuon process \cite{Berger:2016inr,Gao:2017kkx} and included in PDF fits \cite{Bailey:2020ooq,NNPDF:2021njg}. A conceptual issue, which first appears at NNLO, is the infrared-unsafety of inclusive charm-quark production \cite{Forte:2010ta}. This arises from diagrams such as the one in figure \ref{fig:nnlo_charm_pair} which gives rise to a logarithmic contribution $\log(Q^2/m_c^2)$. In fully inclusive DIS, with no requirement of a charm quark in the final state, these logarithtms would be cancelled by virtual corrections such as the one arising from the diagram in figure \ref{fig:nnlo_rv_charm_loop}.

Our generic approach, laid out in refs.~\cite{Helenius:2024fow,Paukkunen:2025kjb}, resolves the aforementioned difficulties at once. The idea is to consider the entire chain of processes that lead to the dimuon production --- the semi-inclusive charm-hadron production followed by its decay. On the one hand, the mass logarithms from diagrams like the one in figure \ref{fig:nnlo_charm_pair} are resummed to the scale dependence of FFs resolving the infrared unsafety discussed above. On the other hand, the non-perturbative decay of charm-hadron into dimuons is treated on rather general grounds by parametrizing the differential decay width and fitting it to the experimental data from $e^+e^-$ collisions \cite{CLEO:2006ivk}. The differential nature of the decay width allows us to naturally evaluate the fiducial cross sections in the decay-muon kinematics, making the separate acceptance correction required in eq.~(\ref{eq:charm_factorization}) an unnecessary concept in our framework.

Higher-order corrections to hard-scattering processes can sometimes entail large corrections due to the opening of new partonic channels which can compensate the suppression induced by the additional powers of the strong coupling $\alpha_s$. Up to NLO, the only allowed initial-state quarks in neutrino scattering are the lower-type quarks (down, strange, bottom). Due to diagrams like the one shown in Figure~\ref{fig:nnlo_square}, the up-type quarks (up, charm) are also possible at NNLO. As the up-quark distribution is enhanced in the valence region, i.e. at large $x$, there could potentially be a significant contribution from the up-quark-initiated channels when approaching this regime.

\section{Theoretical framework}
\label{sec:nnlo}

Consider a neutrino $\nu$ with energy $E_\nu$ incident on a nucleon $N$ at rest. The inclusive cross section for the dimuon final state $\nu N\to \mu\mu X$ with an energy cut $E_\mu^{\min}$ imposed on the muon originating from the decays of charm-hadrons $h$ can be written in the narrow-width approximation as \cite{Helenius:2024fow},
\begin{equation}
    \frac{\dd\sigma(\nu N\to \mu\mu X)}{\dd x \, \dd y}=\sum_h\int \dd z \, \frac{\dd\sigma(\nu N\to \mu hX)}{\dd x \, \dd y \, \dd z} B_{h\to \mu}(E_h=zyE_\nu, E_\mu^{\min}).
    \label{eq:master}
\end{equation}
Here, $\nu N\to \mu hX$ refers to the underlying SIDIS process $\nu_\mu(k) + N(P_N)\to \mu(k')+h(P_h)+X$, whose cross section is of the standard form,
\begin{equation}
\label{eq:sidis_hadron_cross_section}
\begin{aligned}
	\frac{\dd\sigma(\nu_\mu N\to \mu h X)}{\dd x \, \dd y \, \dd z}&= \frac{G_F^2M_W^4}{\left(Q^2+M_W^2\right)^2}
  \frac{Q^2}{2\pi xy}
 \\ &\phantom{=} \ \times \bigg[xy^2 F_1(x, z, Q^2)+\left(1-y-
 \frac{x^2y^2M^2}{Q^2}
 \right)F_2(x, z, Q^2) \\ &\phantom{= \times\bigg[} \ \pm xy\left(1-\frac{y}{2}\right)F_3(x, z, Q^2)\bigg],
\end{aligned}
\end{equation}
where $M$ is the nucleon mass, $G_F\approx 1.166379 \times 10^{-5} \, {\rm GeV}^{-2}$ the Fermi constant, and $M_W\approx 80.377 \, {\rm GeV}$ the $W$-boson mass. The kinematic variables are the usual ones,
\begin{equation}
\label{eq:sidis_kinematics}
\begin{aligned}
Q^2&=-q^2 =- (k-k')^2 \,, \\
x&=\frac{Q^2}{2P_N\cdot q} \,, \\
y&=\frac{P_N\cdot q}{P_N\cdot k} \,, \\
z&=\frac{P_N\cdot P_h}{P_N\cdot q} \,.
\end{aligned}
\end{equation}
The plus sign in the $F_3$ structure function corresponds to neutrino scattering and the minus sign to antineutrino scattering. The function $B_{h \to \mu}$ appearing in eq.~(\ref{eq:master}) is an energy-dependent branching fraction,
\begin{align}
B_{h \to \mu}(E_h, E_\mu^{\min})
& =
\frac{1}{\Gamma_h^{\rm tot}}
\int \dd^3 \mathbf p_\mu \frac{\dd^3\Gamma_{h \to \mu}}{\dd^3\mathbf p_\mu} \Bigg|_{E_\mu>E_\mu^{\min}},
\end{align}
where $\Gamma_h^{\text{tot}}$ is the total decay width of the hadron $h$ and the integration is over the decay-muon momentum $\mathbf p_\mu$ with the muon energy $E_\mu$ restricted above the cut $E_\mu^{\min}$ as typically required in experimental analyses. The generic form of the differential decay width, dictated by Lorentz invariance, can be written as,
\begin{align}
\frac{\dd^3\Gamma_{h \to \mu}}{\dd^3 \mathbf p_\mu}
 = \frac{d_{h \to \mu}(w)}{2m_h E_\mu} \,, \quad w = \frac{p_\mu \cdot P_h}{m_h^2} \nonumber \,,
\end{align}
where $m_h$ is the mass of $h$. The Lorentz-invariant decay function $d_{h \rightarrow \mu}(w)$ has been parametrized and fitted \cite{Helenius:2024fow} to data from $e^+e^-$ collisions \cite{CLEO:2006ivk}. The strucure functions appearing in eq.~(\ref{eq:sidis_hadron_cross_section}) are generic convolutions between PDFs $f_i$, hard coefficient functions $C_{ij}$, and FFs $D_i$,
\begin{equation}
\label{eq:convolution}
F_i(x,z,Q^2) = \sum_{ij} \int_{\chi_n}^1 \frac{\dd\xi}{\xi}\int_z^1 \frac{\dd\zeta}{\zeta} f_i
\left(\frac{\chi_n}{\xi},\fact^2 \right)
 C_{ij}(\xi, \zeta, Q^2, \renorm^2, \fact^2, \frag^2)D_j
 \left(\frac{z}{\zeta},\frag^2 \right) \,,
\end{equation}
where $\renorm^2$ is the renormalization scale, $\fact^2$ the factorization scale, and $\frag^2$ the fragmentation scale. By default, we take $\renorm^2 = \fact^2 = \frag^2 = \mu_0^2=Q^2+m_c^2$. The slow-rescaling variable $\chi_n$,
\begin{equation}
    \chi_n\equiv x\left[1+\frac{(nm_c)^2}{Q^2}\right],
\end{equation}
where $n$ is the number of final-state charm quarks, accounts for the additional kinematical restriction set by the charm-quark masses --- see ref.~\cite{Paukkunen:2025kjb} for more details. The coefficient functions $C_{ij}$ are not unique but they depend on the chosen GM-VFNS variant. Up to NLO, $\mathcal{O}(\alpha_s)$, our calculation here takes them in the so-called SACOT-$\chi$ scheme \cite{Olness:1987ep,Aivazis:1993kh,Aivazis:1993pi,Kretzer:1997pd,Collins:1998rz,Kramer:2000hn,Tung:2001mv,Kretzer:2003it,Guzzi:2011ew,Gao:2021fle,Risse:2025smp}, explained thoroughly in ref.~\cite{Paukkunen:2025kjb}, which reduces to the usual zero-mass $\msbar$ scheme in the limit $Q^2 \to \infty$. The new ingredients here are the NNLO corrections for which we use the massless coefficient functions provided in ref.~\cite{Bonino:2025qta}. This effectively means that we implement them in what can be called the slow-rescaling (SR), or intermediate-mass (IM), approximation \cite{Nadolsky:2009ge}. As we learned in ref.~\cite{Paukkunen:2025kjb}, the kinematic mass corrections that are effectively induced by the slow-rescaling variable, are clearly the dominant ones up to NLO, apart from very low values of $Q^2 \sim m_c^2$ which are always exluded from global PDF fits. Thus, it is reasonable to expect that in the relevant region of $Q^2$ our framework should capture the dominant NNLO corrections.

Along with the size of the perturbative corrections, the dependence of the cross sections on the renormalization, factorization, and fragmentation scales is typically taken as a measure of perturbative convergence. To faithfully study this aspect in our setup, we should also use PDFs and FFs that respect the DGLAP evolution with NNLO splitting functions. For the PDF, we use the \texttt{TUJU21} set \cite{Helenius:2021tof} of nuclear PDFs (iron, $A=56$) which also dictates our choice for the charm-quark mass $m_c=1.43 \, \mathrm{GeV}$. Unfortunately, no applicable NNLO FFs currently exist.\footnote{The initial scale of the $D^0$ and $D^+$ fit in ref.~\cite{Salajegheh:2019nea} is above the bottom-mass threshold, which is too large for our purposes.} In our preceding studies \cite{Helenius:2024fow,Paukkunen:2025kjb} we have used the GM-VFNS NLO sets \texttt{kkks08} (specifically, the OPAL set) \cite{Kneesch:2007ey} for $D^0$ and $D^+$ and \texttt{bkk05} \cite{Kniehl:2006mw} for $D_s$ and $\Lambda_c^+$. In lieu of a proper NNLO fit, we take the analytic NLO parametrizations of these analyses and simply redo the evolution at NNLO using the \texttt{EKO} framework \cite{Candido:2022tld,barontini_2026_18493684}. It is important to note that simply redoing the evolution does not constitute a proper NNLO analysis, but should be sufficient for studying the scale dependence. An additional complication at NNLO is the need for matching conditions in PDF and FF DGLAP evolution when crossing either the charm- or bottom-mass threshold, where the number of active quark flavors $N_f$ changes. Unlike at lower orders, the NNLO coefficient functions depend explicitly on $N_f$, making them discontinuous at the thresholds. This discontinuity should then be compensated in the evolution of the strong coupling $\alpha_s$, PDF, and FF, which also turns these objects discontinuous. In the end, all these discontinuities should cancel so that the cross sections are continuous, though the cancellation is of perturbative nature and therefore numerically incomplete. However, the time-like matching conditions are not fully known at NNLO \cite{Biello:2024zti}. Therefore, they are not implemented in \texttt{EKO} and are not included in our NNLO FFs. As a consequence, the discontinuity at the bottom threshold can still get smaller once the FF matching conditions have been implemented.

\section{Results}

\subsection{NNLO corrections and partonic channels}

\begin{figure}
    \centering
    \includegraphics[width=0.49\linewidth]{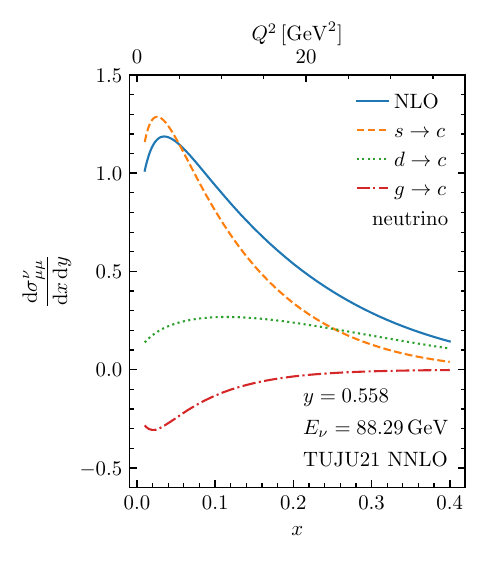}
    \includegraphics[width=0.49\linewidth]{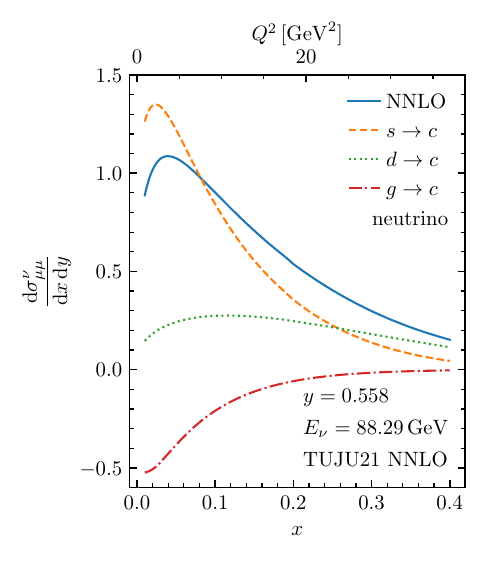}
     \includegraphics[width=0.49\linewidth]{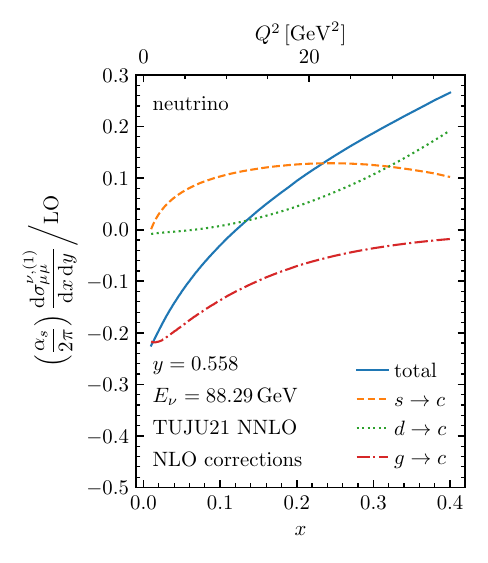}
    \includegraphics[width=0.49\linewidth]{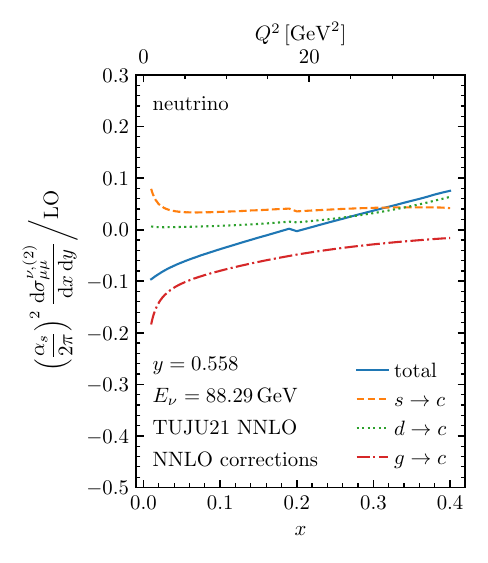}
   \caption{Dimuon cross sections in neutrino-nucleus DIS at NLO (left) and NNLO (right) together with the leading-channel contributions. The upper panels show the full cross sections, while the lower panels show the corresponding perturbative corrections normalized to the full LO cross sections. All contributions are computed with NNLO PDF and FFs, as described in section~\ref{sec:nnlo}.}
    \label{fig:neutrino_leading_channels}
\end{figure}

\begin{figure}
    \centering
    \includegraphics[width=0.49\linewidth]{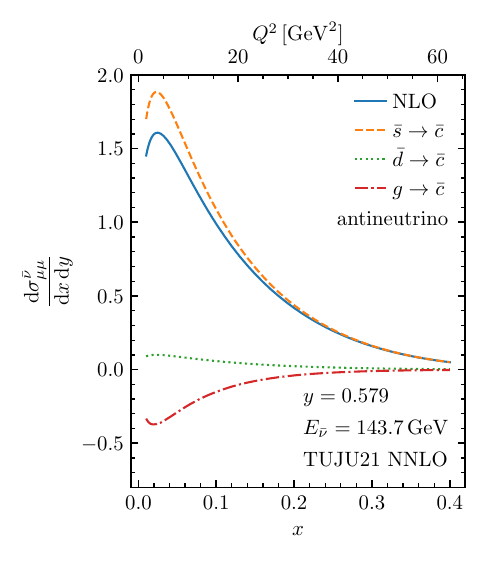}
    \includegraphics[width=0.49\linewidth]{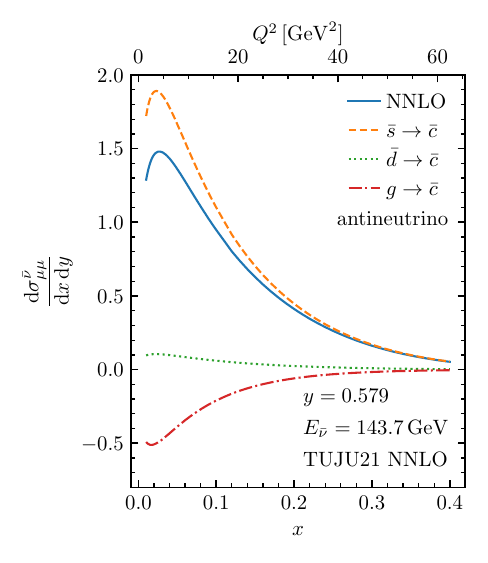}
    \includegraphics[width=0.49\linewidth]{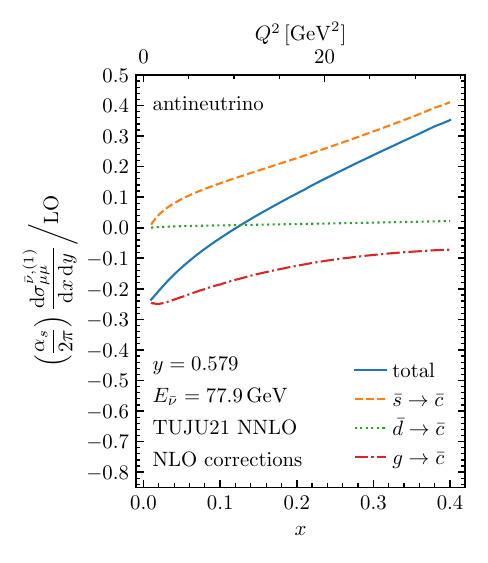}
    \includegraphics[width=0.49\linewidth]{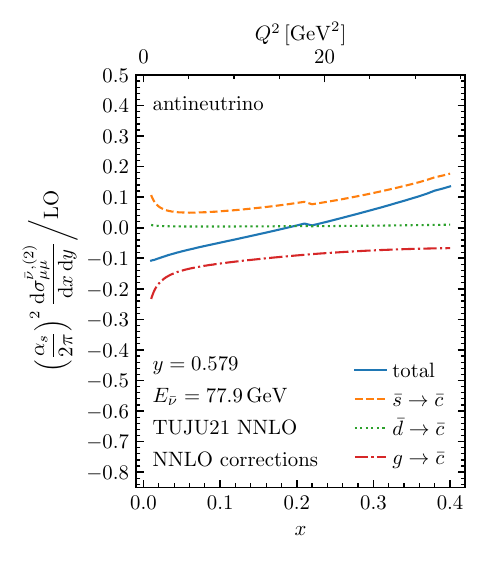}
    \caption{Same as figure \ref{fig:neutrino_leading_channels}, but for antineutrino scattering.}
    \label{fig:antineutrino_leading_channels}
\end{figure}

To understand the impact of NNLO coefficient functions, the upper panels of figures \ref{fig:neutrino_leading_channels} and \ref{fig:antineutrino_leading_channels} display the neutrino and antineutrino cross sections calculated without (left) and with (right) the NNLO corrections. The kinematics correspond to those of the NuTeV experiment. In both cases, the effect of the NNLO terms is to somewhat decrease the cross sections at small values of $x$, while at large $x$ the cross sections grow. Qualitatively similar behaviour has been observed in inclusive charm production \cite{Gao:2017kkx}. The overall effect is, however, rather mild. While the exact numbers of course depend on the PDFs and the kinematics, the trend itself holds generally in the entire kinematic region covered by the existing NuTeV and CCFR measurements. The upper panels of figures \ref{fig:neutrino_leading_channels} and \ref{fig:antineutrino_leading_channels} also show the contributions of the most relevant partonic channels. For neutrino scattering these are $s\to c$, $d\to c$, and $g\to c$, i.e. the ones that are initiated by a down quark, strange quark, or a gluon, and in which the fragmenting parton is the charm quark. Of the quark-initiated channels, the strange-initiated channel is typically more important due to the hierarchy in the Cabibbo-Kobayashi-Maskawa (CKM) matrix elements. This, of course, is the reason why dimuon production is a good constraint for the strange-quark PDFs. Only at very large $x$ does the down-initiated contribution become larger than the strage-initiated one due to the existence of the down-valence PDF in comparison to the strange-quark PDF with only sea-like content. No analogous valence distribution exists for $\bar d$, and therefore antineutrino scattering is entirely dominated by the $\bar s\to \bar c$ and $g\to \bar c$ channels. The lower panels of figures \ref{fig:neutrino_leading_channels} and \ref{fig:antineutrino_leading_channels} compare the NLO, i.e. $\mathcal{O}(\alpha_s)$ (left), and NNLO, i.e. $\mathcal{O}(\alpha_s^2)$ (right), contributions normalized to the full LO cross section. The overall NNLO corrections are seen to be systematically smaller than the NLO ones which indicates a convergence of the perturbative series. In the NNLO corrections the expected discontinuity at the bottom-quark threshold discussed in section~\ref{sec:nnlo} is visible, but in the full cross sections (the upper panels) this small discontinuity can hardly be seen. We find that the NNLO corrections to the gluon-induced channels are always negative while the quark-induced ones are usually positive, and the two tend to compensate each other in the sum. However, it should be noted that only the overall correction is physically relevant as the split to the gluon- and quark-initiated contributions depends heavily on the adopted factorization scheme, and the negativity of the gluon contributions is very typical for the $\msbar$ scheme \cite{Candido:2020yat}. Channels involving the fragmentation of a parton other than the charm quark do not have any meaningful contribution as their charmed-meson fragmentation functions are negligble due to the additional splittings needed to produce a charmed final state. The $b$-quark FF is larger, but the associated CKM-matrix elements quench these contributions.

\begin{figure}
    \centering
    \includegraphics[width=0.49\linewidth]{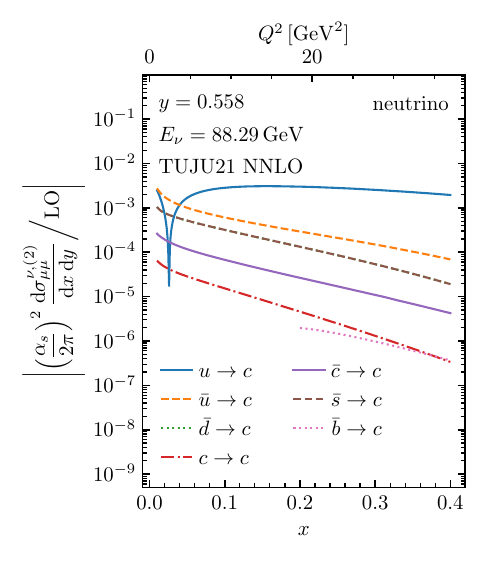}
    \includegraphics[width=0.49\linewidth]{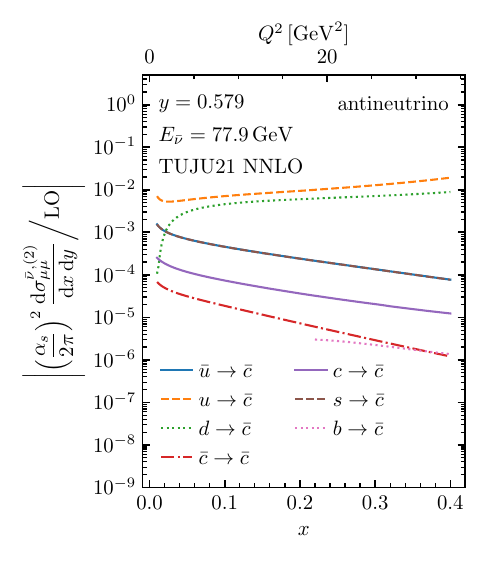}
    \caption{Subleading NNLO corrections to the dimuon cross section in neutrino (left) and antineutrino (right) scattering, normalized to the full LO cross section. The corrections are computed with NNLO PDF and FFs, as described in section~\ref{sec:nnlo}.}
    \label{fig:subleading_channels}
\end{figure}

As discussed in section~\ref{sec:Intro}, new and potentially relevant channels open up at NNLO. These contributions are shown in figure \ref{fig:subleading_channels}, which demonstrates that these new channels are, at the end, not very signficant. For neutrino scattering, the most significant is $u\to c$, which is to be expected because of the large valence contribution from the up-quark distribution. Nevertheless, the contribution from $u\to c$ is still two orders of magnitude smaller than from the pre-existing channels. It should be noted that the contribution from $u\to c$ becomes negative at small $x$, which appears as a kink in the logarithmic plot. For antineutrino scattering, the channels $d\to \bar c$ and $u\to \bar c$ are relatively more important than in the neutrino case as the NNLO is the first order at which the valence-quark PDFs begin to contribute. The contribution from $d\to \bar c$ is comparable to that of $\bar d\to \bar c$ in figure \ref{fig:antineutrino_leading_channels}, with the contribution from $u\to\bar c$ being even larger. While the corrections from these channels are still clearly subleading in comparison to $\bar s\to\bar c$ and $g\to \bar c$, the difference is now one order of magnitude instead of two. It should be noted that in figure \ref{fig:subleading_channels}, the contributions from the $\bar u\to \bar c$ and $s\to \bar c$ channels are identical. This is because these channels use the same coefficient function, and because of the constraint $\bar s=s=\bar d=\bar u$ used in the \texttt{TUJU21} parametrization.

\subsection{Scale uncertainties and comparison with NuTeV data}

\begin{figure}
    \centering
    \includegraphics[width=\linewidth]{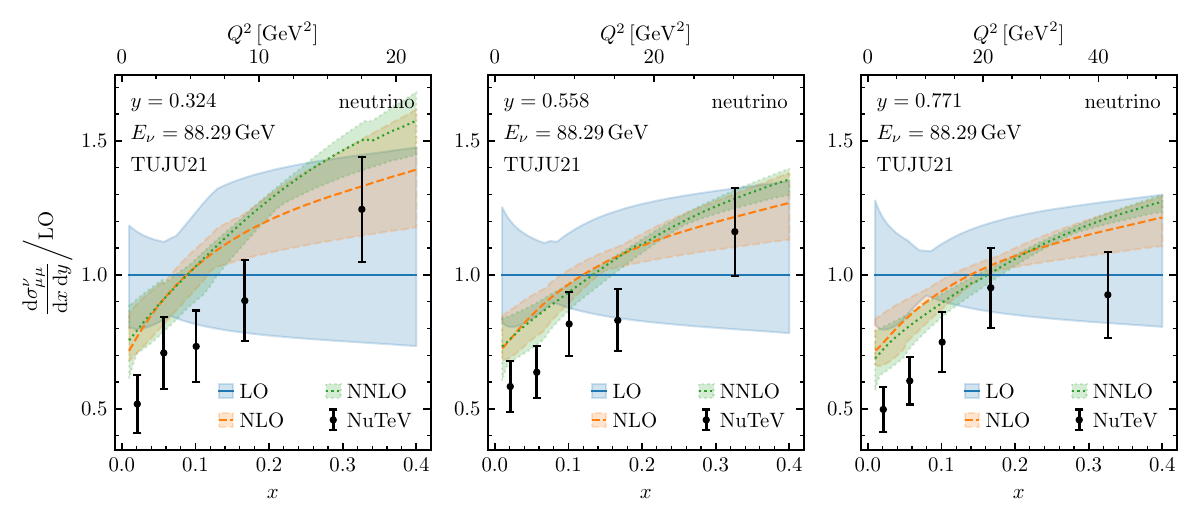}
    \includegraphics[width=\linewidth]{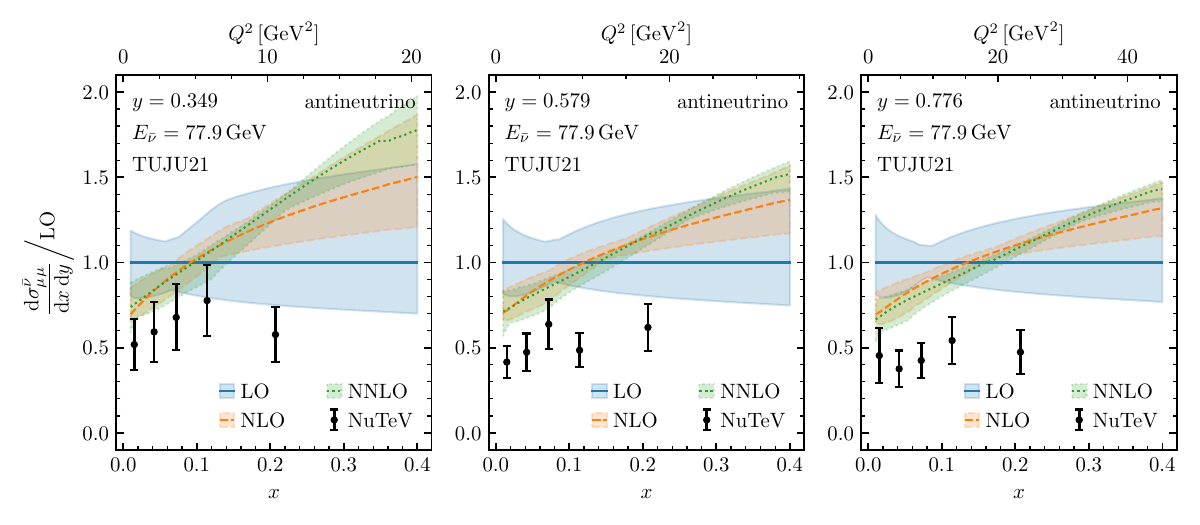}
    \caption{17-point scale variations of the dimuon cross section in neutrino (top) and antineutrino (bottom) scattering, normalized to the LO cross section. The LO and NLO contributions are computed with NLO PDF and FFs, while the NNLO contributions are computed with NNLO PDF and FFs, as described in section~\ref{sec:nnlo}.}
    \label{fig:neutrino_scales17p_tuju21}
\end{figure}

\begin{figure}
    \centering
    \includegraphics[width=\linewidth]{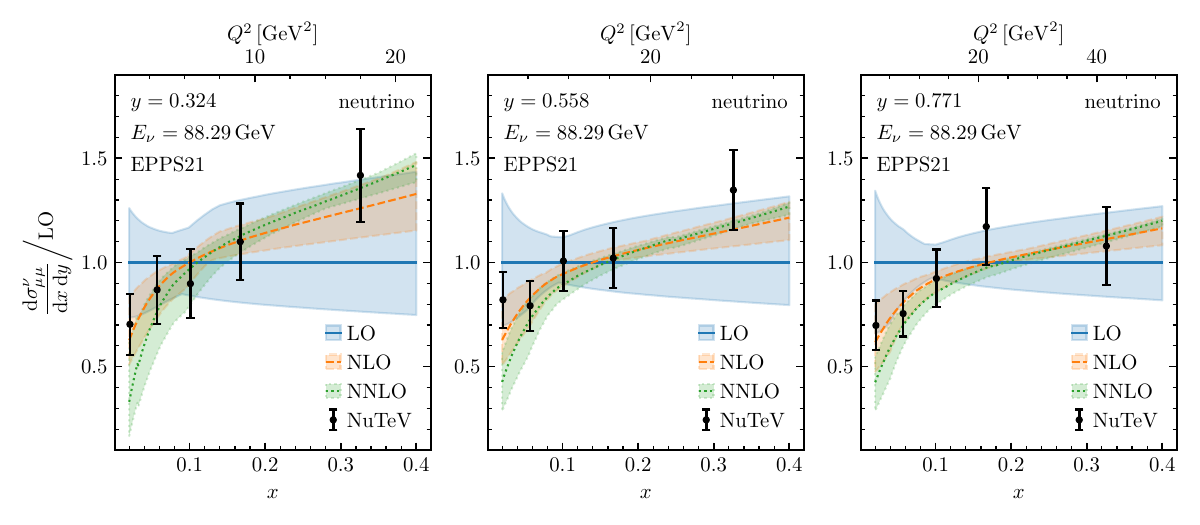}
    \includegraphics[width=\linewidth]{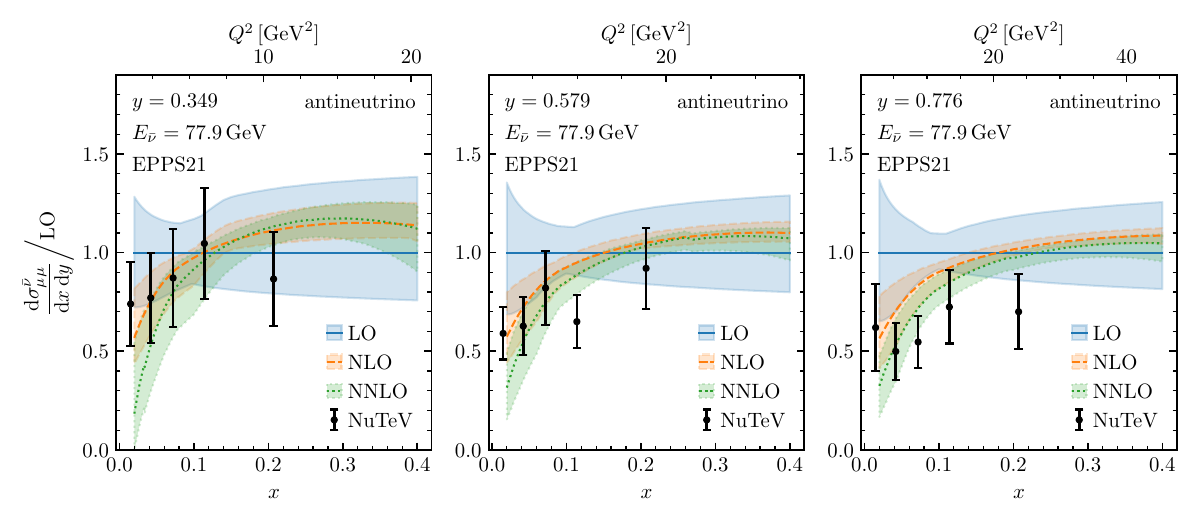}
    \caption{17-point scale variations of the dimuon cross section in neutrino (top) and antineutrino (bottom) scattering, normalized to the LO cross section. All contributions are computed with the NLO \texttt{EPPS21} PDF set and NLO FFs. The theoretical results are compared with NuTeV data.}
    \label{fig:neutrino_scales17p_epps21}
\end{figure}

Figure \ref{fig:neutrino_scales17p_tuju21} shows the 17-point scale variations of the neutrino and antineutrino dimuon cross section at LO, NLO, and NNLO in the same kinematical setup as in the previous plots. We also show the corresponding experimental data from NuTeV \cite{NuTeV:2007uwm}. The LO and NLO results are computed with the \texttt{TUJU21 NLO} PDF and the original \texttt{kkks08} and \texttt{bkk05} FFs, while the NNLO contribution is computed with the \texttt{TUJU21 NNLO} PDF and the NNLO FFs described in section~\ref{sec:nnlo}. The 17-point scale variations consist of variations of the form $\mu_i^2=\max\{k^2\mu_0^2, m_c^2\}$, $k\in \left\{\frac{1}{2}, 1, 2\right\}$, for the renormalization scale $\renorm^2$, factorization scale $\fact^2$, and fragmentation scale $\frag^2$, with the condition $\frac{1}{2}\renorm\leq \fact, \frag\leq 2\renorm$. The LO scale variations are always clearly larger than the NLO or NNLO ones. At small values of $x$ the NNLO scale variations are similar to the NLO ones but as $x$ grows, the NNLO scale variations start to diminish, indicating a better perturbative convergence at larger $x$. This is to be expected as larger $x$ corresponds to larger $Q^2$ and therefore to a smaller expansion parameter $\alpha_s$. One could speculate that the similarity of the NLO and NNLO scale varations at small $x$ could be related to the SR/IM approximation we make in including the NNLO terms. However, at the NLO level we did not observe a significant difference in the scale uncertainties between the SR/IM approximation and the full GM-VFNS calculation \cite{Paukkunen:2025kjb}. The NNLO results always land within the NLO scale-variation bands, so in this case the NLO scale variations gave a reasonable estimate on the size of the missing higher-order corrections. In comparison to the NuTeV data, at small $x$ where the NNLO calculation does not significantly reduce the scale variations, the experimental errors remain comparable to the scale uncertainties. At larger $x$, however, the experimental errors are notably bigger than the NNLO scale uncertainties. Figure \ref{fig:neutrino_scales17p_tuju21} indicates that using \texttt{TUJU21} PDFs for the target nucleon yields on average a 20--30 \% larger cross section compared to the NuTeV data. The mismatch is more pronounced in case of antineutrino data where valence-$d$-originating contributions are negligible. This can be traced to the unsuppressed strangeness in the \texttt{TUJU21} analysis, which assumes $\bar s=s=\bar d=\bar u$ at the parametrization scale. This also concretely demonstrates what we wrote earlier in section~\ref{sec:Intro} about dimuon data preferring a suppressed strangeness and once more highlights the constraints these dimuon data can provide. To this end, figure~\ref{fig:neutrino_scales17p_epps21} presents the same comparison as figure~\ref{fig:neutrino_scales17p_tuju21}, but this time by using the NLO \texttt{EPPS21} nuclear PDF set \cite{Eskola:2021nhw}. This set of PDFs is based on the \texttt{CT18ANLO} set of proton PDFs \cite{Hou:2019efy}, in which the strangeness suppression is a compromise between what these dimuon data --- within a FFNS NLO calculation calculation using the factorized approximation eq.~(\ref{eq:charm_factorization}) --- and the LHC $W$ and $Z$ data prefer. There is now clearly a better correpondence to the shown NuTeV data.

\section{Conclusion and outlook}
\label{sec:conclusion}

In this paper, we presented an NNLO-improved version of our NLO-level GM-VFNS implementation \cite{Helenius:2024fow,Paukkunen:2025kjb} of dimuon production in neutrino-nucleus DIS, based on the zero-mass NNLO SIDIS coefficient functions calculated in ref.~\cite{Bonino:2025qta}. Despite using the zero-mass coefficient functions, we properly took into account the mass dependence that is of kinematic origin and which is arguably the leading mass effect at $Q^2 \gtrsim 4 \, \mathrm{GeV}^2$. We studied the impact the NNLO corrections have on the cross sections at kinematics corresponding to available experimental data and charted the expected size of still higher-order corrections by standard scale variations. In general, the NNLO corrections were found to decrease the cross sections at small $x$ while an enhancement was observed at large $x$. This behaviour appears to decrease the known tension between the strongly-suppressed strange sea traditionally preferred by dimuon data, and the less suppressed strange sea favored by LHC $W/Z$ data at small values of $x$. We saw that the NNLO calculation reduced the scale uncertainties significantly at $x\gtrsim 0.2$. Below that, the NNLO scale variations remained on par with the NLO variations, indicating slower perturbative convergence at small $x$. However, by studying the NLO and NNLO corrections separately, we found that the NNLO corrections are still generally smaller than the NLO corrections, particularly at larger $x$. 

We also considered the impact of new channels that open at NNLO. Most of these are, however, completely negligble for dimuon production. Only the channels initiated by the up or down quark, both of which have a large valence contribution, had a discernable contribution in antineutrino scattering. In the end, however, the bulk of the NNLO contributions arose from corrections to channels already available at NLO. As no NNLO FFs that are applicable to our calculation currently exist, we had to resort to approximative NNLO FFs based on pre-existing NLO fits. A modern NNLO FF fit with proper uncertainty analysis would be beneficial to study the impact and uncertainties associated with the charm-quark fragmentation when using dimuon data in global fits of PDFs. While we expect the NNLO-level mass corrections not included in the presented calculation to be subleading in the $Q^2 \gtrsim 4 \, \mathrm{GeV}^2$ region, a full GM-VFNS calculation would still be needed to reduce the remaining dependence on the adopted GM-VFNS variant. In principle, the FFNS NNLO coefficient functions are already known \cite{Gao:2017kkx}, making this a feasible future extension. Other ways to improve the setup presented here is to consider radiative electroweak and and target-mass corrections, both of which are known to carry importance in neutrino DIS \cite{Paukkunen:2010hb}.

The code used for the numerical results presented in this paper will be available from ref.~\cite{yrjanheikki_2026_18890044} and will also include precomputed interpolation tables enabling an efficient use of our results in global fits of PDFs.

\section*{Acknowledgements}

We thank F. Hekhorn for the help with the \texttt{EKO} package. We acknowledge the financing from the Magnus Ehrnrooth foundation (S.Y.), the Research Council of Finland Project No.~361179 (I.H.), and the Center of Excellence in Quark Matter of the Research Council of Finland, project 364194. We acknowledge grants of computer capacity from the Finnish IT Center for Science (CSC), under the project jyy2580.

\bibliographystyle{JHEP}
\bibliography{refs.bib}

\end{document}